\documentclass[letterpaper]{article} 
\usepackage{aaai2026}  
\usepackage{times}  
\usepackage{helvet}  
\usepackage{courier}  
\usepackage[hyphens]{url}  
\usepackage{graphicx} 
\usepackage{natbib}  
\usepackage{caption} 
\usepackage{algorithm}
\usepackage{algorithmic}

\usepackage{newfloat}
\usepackage{listings}
\DeclareCaptionStyle{ruled}{labelfont=normalfont,labelsep=colon,strut=off} 
\floatstyle{ruled}
\newfloat{listing}{tb}{lst}{}
\floatname{listing}{Listing}
\usepackage[colorlinks=true, linkcolor=black, urlcolor=blue, citecolor=black]{hyperref}
\usepackage{hyperref}
\usepackage{xcolor}  
\title{README: Robust Error-Aware Digital Signature Framework \\ via Deep Watermarking Model}
\author{
Hyunwook Choi$^{1}$ \qquad Sangyun Won$^{1}$ \qquad Daeyeon Hwang$^{1}$ \qquad Junhyeok Choi$^{1}$ \\[5pt]
$^{1}$Korea University\\[5pt]
{\tt \href{https://readme-2025.github.io}{https://readme-2025.github.io}}
}

\usepackage{bibentry}
\usepackage{multirow}
\usepackage{booktabs}
\usepackage{amsmath}
\usepackage{amsfonts}
\nocopyright

\begin{document}

\maketitle

\begin{abstract}
Deep learning-based watermarking has emerged as a promising solution for robust image authentication and protection. However, existing models are limited by low embedding capacity and vulnerability to bit-level errors, making them unsuitable for cryptographic applications such as digital signatures, which require over 2048 bits of error-free data. In this paper, we propose \textbf{README} (\textbf{R}obust \textbf{E}rror-\textbf{A}ware \textbf{D}igital Signature via Deep Water\textbf{M}arking Mod\textbf{E}l), a novel framework that enables robust, verifiable, and error-tolerant digital signatures within images. Our method combines a simple yet effective cropping-based capacity scaling mechanism with \textbf{ERPA} (\textbf{ER}ror \textbf{PA}inting Module), a lightweight error correction module designed to localize and correct bit errors using Distinct Circular Subsum Sequences (DCSS). Without requiring any fine-tuning of existing pretrained watermarking models, README significantly boosts the zero-bit-error image rate (Z.B.I.R) from 1.2\% to 86.3\% when embedding 2048-bit digital signatures into a single image, even under real-world distortions. Moreover, our use of perceptual hash-based signature verification ensures public verifiability and robustness against tampering. The proposed framework unlocks a new class of high-assurance applications for deep watermarking, bridging the gap between signal-level watermarking and cryptographic security.
\end{abstract}
\section{Introduction}

Recent advances in deep learning-based watermarking have attracted significant attention as a promising approach for image protection and authentication. Compared to traditional signal processing approaches~\cite{413536, liu2002svd, 1221662, 1327270}, deep learning watermarking models~\cite{zhu2018hidden, tancik2020stegastamp, jia2021mbrs, zhang2024editguard} offer greater capacity, robustness, and flexibility, enabling a wide range of applications. Recent works have demonstrated high fidelity in reconstructing embedded information while maintaining strong resistance to various image distortions such as additive noise~\cite{zhu2018hidden}, JPEG compression~\cite{jia2021mbrs}, physical distortion~\cite{tancik2020stegastamp} and editing~\cite{zhang2024editguard}. As a result, deep learning watermarking has evolved from a simple information embedding tool into a robust framework capable of addressing more sophisticated security requirements.

However, most existing deep watermarking methods face a common limitation: their embedding capacity is relatively low. Prior works~\cite{zhu2018hidden, tancik2020stegastamp, jia2021mbrs, zhang2024editguard} typically encode around 30 to 64 bits of information, which may be sufficient for copyright tagging or simple identifiers, but falls short of the capacity required for practical security applications. Digital signatures~\cite{1055638, 10.1145/359340.359342, ECDSA} require at least 2048 bits of reliably retrievable data. Even if such high capacity is hypothetically achievable, the challenge of ensuring zero-bit errors becomes increasingly difficult under real-world distortions. Thus, achieving robust, publicly-detectable, and unforgeable watermarking remains extremely challenging~\cite{fairoze2025difficulty}.

To address this challenge, we propose a novel method that preserves the robustness of existing deep watermarking models while enabling error-aware digital signatures. Our method introduces a simple cropping-based strategy to increase embedding capacity and a lightweight module, \textbf{ERPA} (\textbf{ER}ror \textbf{PA}inting Module), to effectively refine errors in the decoded bit sequences. For our ERPA Encoder, we also define a new permutation of 64-bit sequences called Distinct Circular Subsum Sequences (DCSS), which are optimized for error localization and correction, and single-layer decoder that can reliably recover these sequences. To evaluate the reliability of bit extraction under distortion, we introduce a new metric, the Zero Bit Error Image Rate (Z.B.I.R), which measures the proportion of images from which all embedded bits are recovered without any error. Our method significantly improves Z.B.I.R to a practically viable level, even under severe distortions.

By combining these two methods, we introduce our framework, \textbf{README} (\textbf{R}obust \textbf{E}rror-\textbf{A}ware \textbf{D}igital Signature via Deep Water\textbf{M}arking Mod\textbf{E}l). This framework presents a new paradigm for enabling digital signatures in images using robust, error-aware deep watermarking. README overcomes the limitations of prior works in terms of both capacity and bit error rate, enabling reliable embedding and extraction of 2048-bit signatures without requiring any finetuning. As a result, the proposed framework significantly improves the practicality and scalability of watermarking systems in security-critical applications.

Our main contributions are as follows:
\begin{itemize}
    \item We introduce a cropping-based capacity enhancement strategy combined with ERPA, an error repair module, enabling robust digital signatures under high-bit embedding settings.
    \item We propose README, a practical framework for robust, publicly-detectable, and unforgeable digital signatures in images, opening the door to real-world deployment of deep watermarking for secure applications.
    \item ERPA operates on top of existing pretrained deep watermarking models without the need for additional fine-tuning, preserving their inherent robustness in our framework, while drastically improving Z.B.I.R. from 1.2\% to 86.3\%.
\end{itemize}

\section{Related Works}

\subsection{Embedding Strategies in Images}

Embedding additional information into images has been widely studied for ownership verification, covert communication, and tamper detection. Early techniques like Least Significant Bit (LSB) substitution~\cite{413536,lsb,4221886,7226122} were simple but fragile against lossy compression and common image manipulations such as resizing or filtering. Simultaneously, transform-domain methods such as Discrete Cosine Transform (DCT)\cite{1327270, blockdct, 10.1007/s11042-018-5913-9, DCT3}, Discrete Wavelet Transform (DWT)\cite{1221662, dwtdctarticle, BABY2015612}, and Singular Value Decomposition (SVD)~\cite{liu2002svd, ARUNKUMAR2019426} were introduced, but remaining vulnerable to more complex distortions.

With the rise of deep learning, end-to-end watermarking frameworks~\cite{baluja2017deep, rahim2018end, zhu2018hidden, tancik2020stegastamp,  jia2021mbrs, zhang2024editguard} have emerged. These methods jointly learn to embed and extract messages via encoder-decoder architectures, achieving a better trade-off between capacity, imperceptibility, and robustness. By simulating distortions during training, deep watermarking models have significantly advanced robustness against attacks including additive noise~\cite{zhu2018hidden}, JPEG compression~\cite{jia2021mbrs}, physical distortion~\cite{tancik2020stegastamp} and editing~\cite{zhang2024editguard}.

\subsection{Cryptography in Data Embedding}

Information hiding has evolved to address confidentiality, integrity, and provenance verification in images. Consequently, embedding cryptographically protected data within steganography has increased, advancing reliable image authentication. Hash functions~\cite{rfc1321, NIST_FIPS180_1} and digital signatures~\cite{1055638, 10.1145/359340.359342, ECDSA} embedded via steganography enable tamper detection and legitimacy verification of images. In particular, digital signatures utilize public-key cryptography to allow verification without secret keys, while also enabling authentication of the sender’s identity through the use of public keys. 

Recent efforts to incorporate cryptographic primitives into images have largely followed traditional embedding strategies~\cite{alhaj2017crypto, raiyan2025screedsolo}, relying on classical techniques such as LSB, DCT, or DWT~\cite{413536, liu2002svd, 1221662, 1327270}. However, with the advancement of deep learning, there has been growing interest in leveraging deep watermarking models to enable more robust cryptographic embedding~\cite{padhi2024deep, fairoze2025difficulty}. Despite their improved resilience to distortions, these models have been shown to suffer from limited capacity and vulnerability to decoding errors, making it difficult to embed large or sensitive cryptographic payloads such as digital signatures~\cite{fairoze2025difficulty}. Consequently, prior work has primarily focused on embedding hash values as a lightweight alternative~\cite{padhi2024deep}. In this paper, we propose a novel deep learning-based framework that overcomes these limitations, enabling the robust embedding of digital signatures with enhanced error tolerance and capacity.
\begin{figure*}[htbp]
    \centering
    \includegraphics[width=\textwidth]{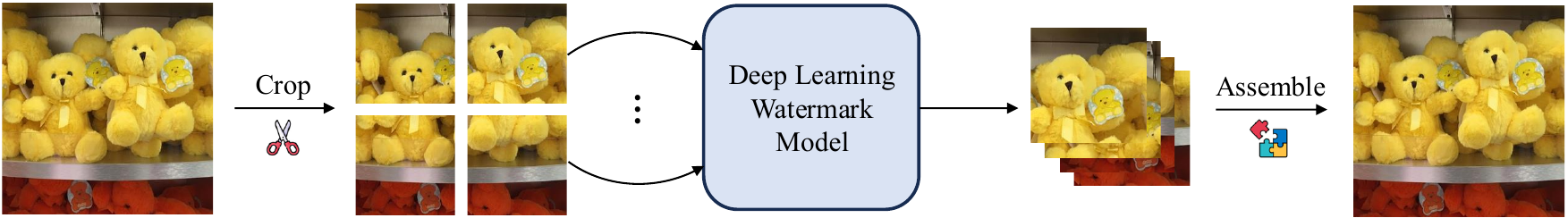}
    \caption{\textbf{Crop-and-then-Watermarking Method.} To increase the embedding capacity while retaining the original watermarking model, the image is first cropped and each crop is processed independently for message embedding. The cropped segments are then reassembled to reconstruct the original image.}
    \label{fig:crop}
\end{figure*}

\section{Methodology}

To overcome the limitations of existing deep watermarking methods in capacity and reliability, our framework combines two key ideas: (1) a cropping-based embedding strategy that linearly scales payload capacity without modifying pretrained models, and (2) an error-aware correction module, ERPA, designed to robustly recover bit-level information under distortion. Together, these components enable our framework README, to reliably embed 2048-bit digital signatures framework in images, laying the foundation for robust, secure, high-capacity watermarking.

\subsection{Cropping-Based Capacity Scaling}

Conventional deep learning-based watermarking models operate on entire images as indivisible units, resulting in limited embedding capacity, typically sufficient only for short identifiers or metadata. This limitation is particularly problematic for cryptographic applications such as RSA~\cite{10.1145/359340.359342} or ECDSA~\cite{ECDSA}, which require the reliable embedding of thousands of bits per image.

To overcome this constraint, as shown in Figure~\ref{fig:crop}, we propose a \textit{crop-and-then-watermark} method that crops the input image into multiple patches and embeds independent bit sequences into each patch. This approach linearly scales the overall embedding capacity with the number of patches, without altering architecture or parameters of the pretrained watermarking model.

Given an input image \(I\), we divide it into \(n\) sub-images \(\{I_1, I_2, \dots, I_n\}\) using a uniform grid. Each sub-image \(I_i\) is processed by an identical instance of a pretrained watermarking network \(W(\cdot)\), which embeds a bit-stream \(m_i\):

\[
I_i' = W(I_i, m_i), \quad i = 1, \dots, n.
\]

The watermarked patches \(\{I_1', \dots, I_n'\}\) are subsequently reassembled to form the final watermarked image:

\[
I^* = \text{Recompose}(\{I_1', \dots, I_n'\}).
\]

Decoding follows the same patch-wise procedure. The watermarked image \(I^*\) is re-cropped identically, and each patch is decoded independently to recover the message \(\tilde{m_i}\). The final decoded message \(\tilde{m}\) is reconstructed via concatenation:

\[
\{\tilde{m_1}, \tilde{m_2}, \dots, \tilde{m_n}\} \rightarrow \tilde{m}.
\]

\begin{figure}[t]
    \centering
    \includegraphics[width=\columnwidth]{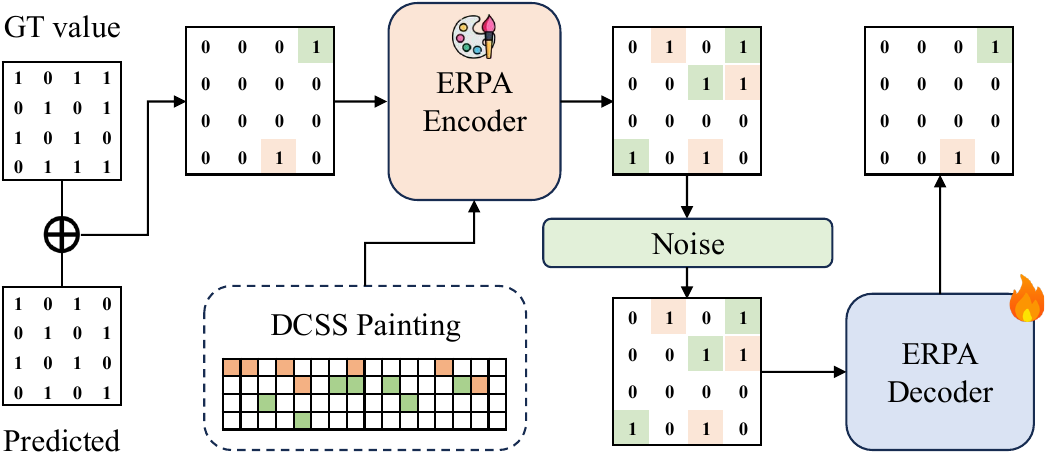}
    \vspace{-15pt}
    \caption{\textbf{ERror PAinting Module (ERPA).} ERPA enhances bit-level watermark reliability through a deterministic encoder and a noise-robust neural decoder. The encoder spatially distributes error vectors using a Distinct Circular Subsum Sequence (DCSS), allowing the decoder to recover the original error pattern even under distortion.}
    \label{fig:erpa}
    \vspace{-10pt}
\end{figure}

\begin{figure*}[htbp]
    \centering
    \includegraphics[width=\textwidth]{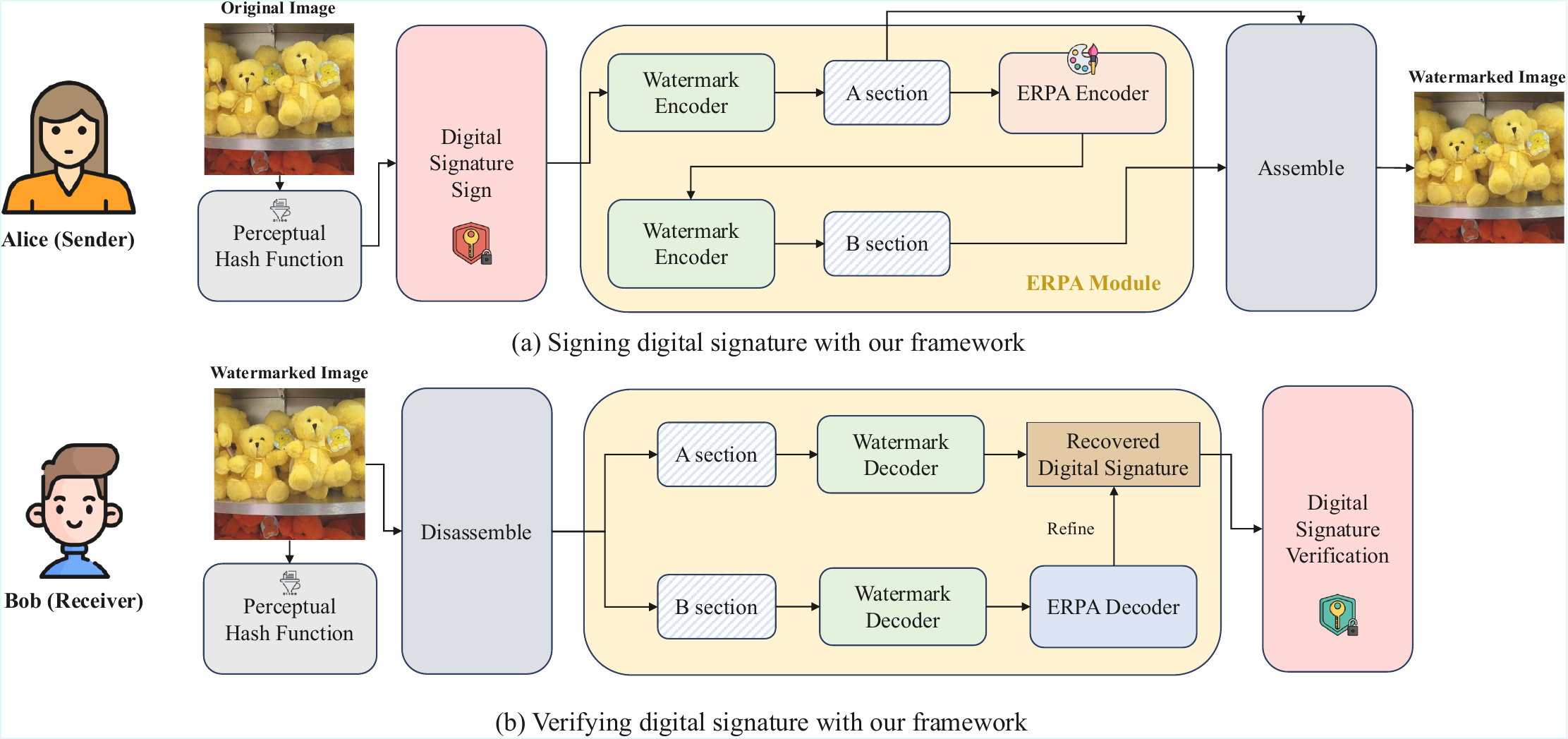}
    \caption{\textbf{Overall pipeline of the proposed README framework.} The framework embeds a perceptual-hash-bound digital signature across cropped image patches, supported by a patch-wise error correction module (ERPA). This enables high-capacity, robust, and verifiable authentication, even under image distortions.}
    \label{fig:framework}
\end{figure*}

\subsection{ERror PAinting Module (ERPA)}

To enable error-aware watermarking suitable for cryptographic use cases, we propose ERPA, a novel mechanism designed to improve the bit-level reliability of watermark decoding under distortion.

\paragraph{Overview.}
Given a message patch \(A\) with ground-truth watermark \(m \in \{0,1\}^n\), and its predicted version \(\tilde{m}\), the bitwise error vector \(e \in \{0,1\}^n\) is computed as:

\[
e = m \oplus \tilde{m}, \quad \text{where } e_i = \begin{cases} 
1 & \text{if } m_i \ne \tilde{m}_i \\
0 & \text{otherwise}
\end{cases}.
\]

This error vector represents the locations of bit mismatches in patch \(A\). We assume that the error pattern within patch \(A\) is deterministic and consistent during decoding. Therefore, instead of discarding this information, we embed it into an error-handling patch \(B\) to support downstream error correction. However, directly embedding \(e\) is vulnerable, as patch \(B\) may itself suffer from bit errors during decoding. To address this, ERPA paints \(e\) into a more robust and spatially redundant representation \(e'\), designed to tolerate corruption. In this context, painting refers to encoding the error vector into spatially distributed offsets to support robust reconstruction.

\paragraph{Deterministic DCSS Painting Encoder.}
A key design goal in painting \(e\) to \(e'\) is to avoid ambiguity when multiple errors occur. To this end, we introduce the \textit{Distinct Circular Subsum Sequences (DCSS)}. We treat the 64-bit vector as a circular array and define a non-overlapping permutation scheme that determines the relative distances over which each error is painted to nearby positions. The error painting offsets, defined by the DCSS distances, are distributed across multiple distinct positions for each bitwise error. This ensures that no two distinct subsets of positions result in the same cumulative offset modulo 64. Formally, for a set of 64 bits DCSS \(S\), condition requires:

\begin{equation}
\begin{array}{c}
    \text{Let } S \subset \mathbb{N} \text{ such that } \sum_{s \in S} s = 64, \\[6pt]
    \forall \, \text{contiguous subsequences } S_1, S_2 \subseteq S, \\[4pt]
    \sum_{k \in S_1} k \not\equiv \sum_{j \in S_2} j \mod 64.
\end{array}
\end{equation}
\vspace{1pt}

This constraint ensures that overlapping shifted positions do not lead to ambiguous collisions during decoding, even under circular wrapping.

We construct a DCSS distance offset set for \(n = 64\) that maximizes the number of elements within a 64-bit space:
\[
S = \{1, 2, 4, 5, 8, 10, 34\}, \quad S_{\text{offset}} = \{0, 1, 3, 7, 12, 20, 30\}.
\]

Each error bit \(e_i = 1\) is painted to the positions defined by:
\[
\text{shift}(S, i) = \left\{ (i + k) \bmod 64 \;\middle|\; k \in S_{\text{offset}} \right\},
\]
and the final encoded sequence \(e'\) is constructed by aggregating all such shifts:
\[
e' = \bigvee_{i=1}^{n} \left( e_i \cdot \text{shift}(S, i) \right).
\]

This DCSS painting ensures high separability between multiple error locations and improves robustness against decoding noise.

\paragraph{Noise-Robust Neural Decoder.}
To recover the original error vector \(e\) from the noisy decoded sequence \(\tilde{e}'\), we employ a lightweight neural decoder. The decoder is trained to invert the spreading and redundancy operations, even in the presence of error corruption.

During training, we simulate conditions by randomly flipping bits in \(e'\) to generate \(\tilde{e}'\), and use supervision from the ground-truth \(e\) to learn a robust mapping:

\[
\hat{e} = \text{Decoder}(\tilde{e}').
\]

This enables the model to generalize across a wide range of distortion levels and error densities. A visualization of the encoding and decoding pipeline is shown in Figure~\ref{fig:erpa}.
Given the extracted watermark \( \tilde{m} \) and the predicted error vector \( \hat{e} \), an estimate of the original watermark \( m \) can be obtained via a bitwise XOR operation:
\[
\hat{m} = \tilde{m} \oplus \hat{e}.
\]

\subsection{Robust Error-Aware Digital Signature Framework}

README, our proposed framework, unifies a cropping-based embedding strategy with the \textit{ERror PAinting Module}. While recent deep learning-based watermarking techniques have significantly advanced in imperceptibility and robustness, they remain insufficient for cryptographic applications due to limited payload capacity and vulnerability to minor image distortions that can lead to critical bit-level errors during extraction. These limitations hinder their use in high-assurance scenarios such as digital signature verification, which require both high capacity and near-zero bit error rates. As shown in Figure~\ref{fig:framework}, our framework addresses these challenges via modular components for capacity scaling, error correction, and perceptual hash-based signature verification.

\subsubsection{Embedding Capacity and Cropping Structure}

Digital signature~\cite{1055638, 10.1145/359340.359342, ECDSA} require payloads of at least 2048 bits, far exceeding the capacity of conventional watermarking models, which typically support 30–64 bits per image.
To address this, we adopt a uniform cropping strategy that partitions an image of size $H \times W$ into an $8 \times 8$ grid, yielding $64$ sub-images of size $(H/8) \times (W/8)$ each. Each sub-image is processed independently using a shared, pre-trained watermarking network, embedding 64 bits per patch. This structure scales the overall capacity to $4096$ bits per image, sufficient to embed both the digital signature and auxiliary error-handling information.

\subsubsection{Robust Error Correction via ERPA}

To ensure reliable extraction under challenging situations, we introduce the \textit{ERror PAinting Module}, a patch-wise redundancy mechanism for correcting bit-level errors. Among the 64 cropped patches, 32 are designated as $A$ patches for embedding the 2048-bit digital signature, and the remaining 32 are assigned as $B$ patches to support error correction for their corresponding $A$ counterparts. The bit error sequence from message patch $A$ is encoded to error-handling patch $B$ using a Distinct Circular Subsum Sequence (DCSS), to minimize overlap and improve the robustness of recovery. The ERPA decoder is pretrained from Bernoulli-distributed bit errors over 64-bit inputs. As a result, it can be directly plugged in without requiring prior knowledge of the exact error distribution.

\subsubsection{Perceptual Hash-Based Signature Verification}

Our framework adopts perceptual hashing (pHash) for signature generation, instead of conventional cryptographic hashes. Standard hash functions (e.g., SHA-256) produce drastically different values under minor changes, making them unsuitable for verifying the watermarked image without the original. Standard cryptographic hashes (e.g., SHA-256) are highly sensitive to minor image distortions, making them unreliable for signature verification when only the watermarked image is available. Perceptual hashes (pHash) remain stable under benign visual changes, making them suitable for image-dependent signatures. The signature is computed as $Sign(sk, \text{pHash(image)})$ and embedded into the image. As pHash values are robust to distortions, the watermarked and original images yield nearly identical results.

This enables verification using only the public key and the watermarked image, without requiring access to the original. Consequently, our framework enables robust and content-aligned authentication suitable for deployment on open platforms where original image recovery is infeasible.

\subsubsection{Security Analysis and Attack Resistance}

Our framework ensures robust security under both encoder-open and encoder-closed threat models, leveraging the proven cryptographic strength of digital signatures and the inherent robustness of deep watermarking.

We assume a standard cryptographic premise: adversaries cannot forge valid signatures without the private key. In the encoder-open setting, attackers may embed arbitrary data or perform replay attacks by extracting and reusing signatures. However, our perceptual-hash-bound signatures inherently tie the signature to specific image content, thereby invalidating verification upon transplantation. Thus, signature reuse for impersonation is effectively prevented.

In the encoder-closed setting, adversaries lack access to the embedding model, making unauthorized watermark insertion practically infeasible, thereby inherently preventing impersonation.

\begin{table*}[t]
    \centering
    \resizebox{\textwidth}{!}{
        \begin{tabular}{l|l|cccc|cc|cc}
            \toprule
            & Methods 
            & Image Size & Crop Size & M.L.(max) & Eval. Payload & PSNR(dB)  & SSIM & Z.B.I.R(\%) & BER(\%) \\
            \midrule
            (I) & Baseline~\cite{jia2021mbrs} & 128x128 & 1×1 & 64 & 64 & 42.67  & 0.9753 & 100.0 & 0.0000  \\
            \midrule
            (II) & (I) + w/ CW (\textbf{Ours}) & 512×512 & 4×4 & 1024 & 512 & \multirow{2}{*}{43.00} & \multirow{2}{*}{0.9633} & 99.9 & 0.0002 \\
            (III) & (II) + w/ ERPA (\textbf{Ours}) & 512×512 & 4×4 & 512 & 512 & & & \textbf{100.0} & \textbf{0.0000} \\
            \midrule
            (IV) & (I) + w/ CW (\textbf{Ours}) & 1024×1024 & 8×8 & \textbf{4096}& 2048 & \multirow{2}{*}{44.62} & \multirow{2}{*}{0.9596} & 99.7 & 0.0001 \\
            (V) & (IV) + w/ ERPA (\textbf{Ours}) & 1024×1024 & 8×8 & \textbf{2048} & 2048 & & & \textbf{100.0} & \textbf{0.0000} \\
            \bottomrule
        \end{tabular}
    }
    \vspace{5pt}
    \caption{
    \textbf{Evaluation of Image Quality and Robustness.}
    We report PSNR, SSIM, BER, and Zero BER Image Rate (Z.B.I.R) to assess both visual fidelity and reliable message recovery. M.L. is the maximum embeddable message length, while Eval. Payload is the actual payload used in evaluation, adjusted to account for ERPA's redundancy. 
    Paired settings use equal Eval. Payload for fair comparison.
    }
    \label{tab:img_qual}
\end{table*}
\begin{table*}[t]
    \centering
    \resizebox{\textwidth}{!}{
        \begin{tabular}{l|l|cccc|cc}
            \toprule
            & Methods 
            & Image Size & Crop Size & M.L.(max) & Eval. Payload 
            & Z.B.I.R(\%) & BER (\%) \\
            \midrule
            (I) & Baseline~\cite{jia2021mbrs} + w/ JPEG ($Q=50$)
                & 128×128 & 1×1 & 64  & 64 
                & 86.6 & 0.3141\\
            \midrule
            (II) & (I) + w/ CW (\textbf{Ours}) 
                 & 512×512 & 4×4 & 1024 & 512
                 & 37.3 & 0.5176  \\
            (III) & (II) + w/ ERPA (\textbf{Ours}) 
                  & 512×512 & 4×4 & 512 & 512 
                  & \textbf{95.2} & \textbf{0.1092} \\
            \midrule
            (IV) & (I) + w/ CW (\textbf{Ours}) 
                 & 1024×1024 & 8×8 & \textbf{4096} & 2048
                 & 1.2 & 0.6679 \\
            (V) & (IV) + w/ ERPA (\textbf{Ours}) 
                & 1024×1024 & 8×8 & \textbf{2048}  & 2048
                & \textbf{86.3} & \textbf{0.1395} \\
            \bottomrule
        \end{tabular}
    }
    \vspace{5pt}
    \caption{\textbf{Quantitative Comparison of Robustness under JPEG Compression.} We report BER and Zero BER Image Rate (Z.B.I.R) under distortion settings. The results highlight the impact of cropping and the effectiveness of ERPA in restoring robustness.}
    \label{tab:jpeg}
\end{table*}

\begin{figure}[htbp]
    \centering
    \includegraphics[width=\columnwidth]{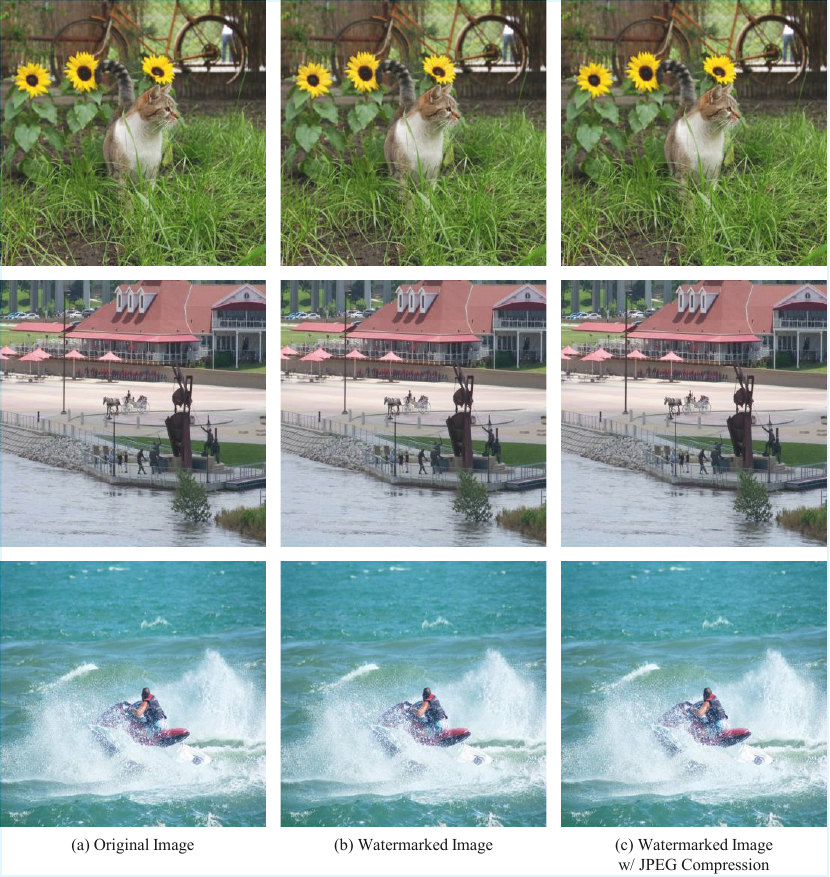}
        \vspace{-15pt}
    \caption{\textbf{Qualitative Results of Watermarked Images.} Visual comparison between original and watermarked images shows that our crop-based method preserves structural details and introduces no perceptible artifacts.}
    \label{fig:watermark}
        \vspace{-10pt}
\end{figure}
\section{Experiments}

\subsection{Implementation Details}

\paragraph{Watermarking Model.}

While our proposed framework is designed to be compatible with a wide range of deep learning-based watermarking architectures, we adopt MBRS~\cite{jia2021mbrs} as our baseline for evaluation. MBRS is a widely used and publicly available watermarking model that supports robust 64-bit message embedding into $128 \times 128$ images using an auto-encoder trained with both simulated and real JPEG compression.

We utilize the official PyTorch implementation of MBRS, employing the publicly released weights optimized for JPEG robustness. This choice serves as a practical benchmark, not a model-specific constraint, allowing us to demonstrate the general applicability of our method under realistic conditions. Importantly, our method does not require any modification to the underlying MBRS architecture, highlighting its plug-and-play compatibility with existing watermarking systems.

\paragraph{Cropping Methods.}

To support high-capacity watermark embedding, particularly for cryptographic use cases, our framework adopts a \textit{crop-and-then-watermark} procedure. Since our baseline enables embedding into $128 \times 128$ images, input images are resized to either $512 \times 512$ or $1024 \times 1024$, and partitioned into regular grids of $4 \times 4$ or $8 \times 8$, resulting in 16 or 64 sub-images. Each sub-image independently carries a 64-bit watermark, scaling the total capacity linearly (up to 4096 bits for $8 \times 8$).

This structure enables reliable embedding of large payloads such as RSA signatures (2048 bits), while leaving room for error correction data. This approach maintains the robustness of the model and localizes potential decoding errors to individual patches, ensuring that errors remain consistent within each patch.

\paragraph{Metrics.}

We evaluate our framework on 1k randomly sampled images from the COCO test dataset~\cite{lin2014microsoft}, focusing on reducing bit errors while maintaining image quality and preserving the robustness of the underlying deep watermarking model.

For image quality, we report average PSNR and SSIM scores. For recovery performance, we use \textit{Bit Error Rate (BER)}. However, since cryptographic signatures require exact bitwise recovery, we additionally introduce the \textit{Zero BER Image Rate (Z.B.I.R)}, the percentage of images with perfectly recovered messages. Z.B.I.R directly reflects the system’s suitability for security-critical applications, where even a single bit error in cropped images is unacceptable.

\paragraph{ERPA Module.}

To improve robustness against bit-level distortions, we introduce the \textit{ERror PAinting Module (ERPA)}, a lightweight error correction mechanism designed to enhance Z.B.I.R without modifying the underlying watermarking model. We empirically found that using 7 offset elements, the maximum permissible in our 64-bit DCSS design, yields the best performance under low-error conditions, and thus adopt this configuration for experiments.

Implemented in PyTorch, the ERPA decoder consists of a single $64 \times 64$ linear layer followed by a sigmoid activation. It is trained independently with a learning rate of $1 \times 10^{-2}$ and batch size of 64, requiring minimal computational overhead. Training is conducted using a Bernoulli error distribution, where each bit is independently flipped with a fixed probability. A fixed permutation is applied to paint error bits, and the decoder is trained to reverse this transformation and recover the original message.

\subsection{Maintaining Visual Quality}

Our cropping-based embedding may introduce local artifacts or discontinuities. To assess the impact on visual fidelity, we compare our method with the MBRS baseline using PSNR and SSIM in Table~\ref{tab:img_qual}, alongside qualitative analysis (Figure~\ref{fig:watermark}). 

Results show preserved visual quality, with no visible artifacts or structural inconsistencies. This confirms that localized embedding does not degrade image quality, even with fine-grained partitioning. However, cropping reduces semantic content per patch, degrading BER and Z.B.I.R even without noise.

To address this, we integrate the ERPA module. Although ERPA reserves half of the redundancy capacity, the total embedding space of 4096 bits in our framework (with $8 \times 8$ cropping) easily accommodates a 2048-bit RSA signature along with correction data. 

To ensure fair comparison with ERPA, we evaluate both ERPA and non-ERPA settings using the same payload length for Z.B.I.R, matching the reduced capacity available when ERPA is applied. With ERPA, the system achieves \textbf{100\%} Z.B.I.R under near-ideal conditions, ensuring robust signature verification without sacrificing visual quality. This demonstrates that our method meets the dual demands of perceptual fidelity and cryptographic reliability, making it suitable for our framework.

\subsection{Maintaining Robustness}

Digital signatures require exact bitwise recovery; thus, we focus on Z.B.I.R as a primary metric. As shown in Table \ref{tab:jpeg}, we evaluate the baseline MBRS model under JPEG compression. Although the average BER seems low (e.g., 0.3141\%), the probability of flawless 64-bit recovery is approximately 86.6\% which is critical considering the bit length involved. Cropping further exacerbates the issue: a $4 \times 4$ grid increased BER to 0.5176 with Z.B.I.R dropping to 37.3\%. In the $8 \times 8$ grid setting, required by our framework, Z.B.I.R is reduced to just 1.2\%. However, ERPA significantly improves robustness. In our README framework, BER dropped from 0.6679 to 0.1395, while Z.B.I.R rose dramatically from \textbf{1.2\%} to \textbf{86.3\%}. 

These improvements demonstrate that ERPA enables cryptographic-level reliability even under distortion, validating its effectiveness as a lightweight, model-agnostic enhancement. These results confirm that the proposed framework enables reliable digital signature embedding while preserving the model's robustness, even under image distortions.

\begin{table}[htbp]
    \centering
    \small
    \setlength{\columnwidth}{4pt}
    \renewcommand{\arraystretch}{1.2}
    \begin{tabular}{c|cccccc}
        \toprule
        \textbf{L} & $p$=0.01 & 0.02 & 0.03 & 0.05 & 0.1 & 0.15\\
        \midrule
        2 & 0.9994 & 0.9982 & 0.9952 & 0.9835 & 0.9555 & 0.9047 \\
        3 & 0.9997 & 0.9987 & 0.9968 & 0.9904 & 0.9587 & \textbf{0.9100} \\
        4 & 0.9997 & 0.9958 & 0.9986 & 0.9943 & \textbf{0.9630} & 0.9038 \\
        5 & 0.9999 & \textbf{0.9999} & 0.9993 & 0.9949 & 0.9624 & 0.9004\\
        6 & 0.9999 & \textbf{0.9999} & 0.9995 & \textbf{0.9968} & 0.9618 & 0.8917 \\
        7 & \textbf{1.0000} & \textbf{0.9999} & \textbf{0.9997} & 0.9966 & 0.9558 & 0.8834 \\
        \bottomrule
    \end{tabular}
    \caption{\textbf{Effect of Painting Density on Bit Error Rate (BER).} BER of the ERPA decoder under Bernoulli($p$) noise as a function of painting redundancy $L$ (i.e., number of DCSS offsets per error bit). Moderate painting (e.g., $L=7$) achieves optimal robustness in practical settings.}
    \label{tab:paint_num}
\end{table}

\begin{table}[htbp]
    \centering
    \resizebox{\columnwidth}{!}{%
    \begin{tabular}{l|cccccc}
        \toprule
         & $p$=0.01 & 0.02 & 0.03 & 0.05 & 0.1 & 0.15 \\
        \midrule
        \textbf{DCSS} & \textbf{1.0000} & \textbf{0.9999} & \textbf{0.9997} & \textbf{0.9966} & \textbf{0.9558} & \textbf{0.8834} \\
        NearBy & 0.9994 & 0.9974 & 0.9936 & 0.9797 & 0.9232 & 0.8573 \\
        \bottomrule
    \end{tabular}
    }
    \caption{\textbf{Impact of Permutation Strategy on BER.} Comparison between DCSS and local (adjacent-offset) painting strategies under Bernoulli($p$) noise, using 7 painted positions. DCSS consistently achieves higher recovery accuracy across all noise levels, owing to its superior spatial separability.}
    
    \label{tab:permutations}
\end{table}

\begin{table}[t]
    \centering
    \resizebox{\columnwidth}{!}{ 
        \begin{tabular}{l|cc}
            \toprule
            Trained Distribution
            & Z.B.I.R(\%) & BER (\%) \\

            \midrule
            Known Exact Error
                & 83.7 & \textbf{0.1145} \\
            Known Error Probability
                & 80.5 & 0.1699 \\
            \midrule
            $Bernoulli(p=0.01)$
                & 76.50 & 0.2391 \\
            $Bernoulli(p=0.05)$
                 & 86.20  & 0.1411 \\
            $Bernoulli(p=0.07)$
                & \textbf{86.30}  & 0.1395 \\
            $Bernoulli(p=0.1)$
               & 86.20 & 0.1409 \\
            \bottomrule
        \end{tabular}
    }
    \caption{\textbf{Impact of Training Distribution on ERPA Decoder Performance.} We compare different training regimes for the ERPA decoder: oracle supervision (exact error), known error probability, and Bernoulli-distributed error modeling. While oracle training achieves the lowest BER, Bernoulli($p=0.07$) training yields the highest Zero BER Image Rate (Z.B.I.R), demonstrating greater robustness under practical conditions.}
    
    \label{tab:train_prob}
\end{table}

\subsection{ERPA Analysis}

We conducted a set of experiments to evaluate the effectiveness of the proposed ERPA module in three key areas: (1) the effect of painting density, (2) the benefit of the DCSS permutation strategy, and (3) the influence of training Bernoulli distribution for error modeling.

\paragraph{Effect of Painting Density}

Table~\ref{tab:paint_num} summarizes how varying the number of painted bits affects decoding performance under different Bernoulli error probabilities \(p\). We adopt the DCSS (Distinct Circular Subsum Sequences) scheme to distribute each error bit across multiple positions while minimizing overlap. In this scheme, each bit with error is painted onto a deterministic set of 7 positions using circular offsets that satisfy the DCSS condition.

Experimental results show a clear trend: when \(p \leq 0.05\), increasing the number of painted bits improves the decoder’s accuracy, as redundancy enhances robustness. However, at higher noise levels, excessive painting leads to performance degradation, likely because the painted positions themselves are increasingly corrupted. Thus, using 7 DCSS-painted offset per error bit offers the best trade-off under practical noise conditions typical of watermarking applications.

\paragraph{Effectiveness of DCSS Painting Strategy}

To assess the benefit of DCSS over simpler alternatives, we compare it with a local painting scheme that selects nearby positions in a 1D message vector. As shown in Table~\ref{tab:permutations}, both methods use 7 painted positions for fair comparison.

DCSS significantly outperforms local painting across all \(p\) values. This is attributed to its high spatial separability and minimal collision. In contrast, local painting is more prone to redundancy overlap and ineffective when errors are distributed non-locally, as is common in distortions like compression or dropout.

These findings highlight that the spatial distribution of redundancy, not just its quantity, is critical for generalizable error correction.

\paragraph{Effectiveness of Bernoulli-Based Training}

To evaluate ERPA under practical conditions, we tested it within README framework, embedding 2048-bit messages. We then compared three training settings for the decoder (Table~\ref{tab:train_prob}):

\begin{itemize}
    \item \textbf{Known Exact Error}: The decoder is trained with full supervision of error locations (oracle).
    \item \textbf{Known Error Probability}: The decoder knows the global error probability density but not specific positions.
    \item \textbf{Bernoulli Error Assumption}: Each bit is assumed to flip independently with probability \(p\).
\end{itemize}

Although the oracle model achieved the lowest Bit Error Rate (BER), it underperformed on the Zero BER Image Rate (Z.B.I.R), a critical metric in digital signature applications requiring perfect bit recovery. In contrast, the Bernoulli-trained decoder achieved the highest Z.B.I.R, indicating better robustness in real-world settings.

These results suggest that while precise error supervision reduces the average BER, it tends to overfit to rare high-error cases, compromising performance under typical low-error conditions. In contrast, Bernoulli-based training is better aligned with typical error distributions, resulting in more reliable full-sequence recovery and thus higher Z.B.I.R.

Furthermore, Bernoulli-based training offers strong practical advantages: it requires no knowledge of error profiles and generalizes across models and distortion types. This makes it well-suited for scalable deployment in secure watermarking systems where cryptographic verification demands strict bitwise fidelity.

\bibliography{main}

\end{document}